# Moment-to-Moment Detection of Internal Thought from Eye Vergence Behaviour

Michae Xuelin Huang[1], Jiajia Li[2], Grace Ngai[2], Hong Va Leong[2], Andreas Bulling[3]

[1]Max Planck Institute of Informatics, Saarland Informatics Campus

[2]Department of Computing, the Hong Kong Polytechnic University

[3]Institute for Visualisation and Interactive Systems, University of Stuttgart, Germany

mxhuang@mpi-inf.mpg.de[1], {csjjli, csngai, cshleong}@comp.polyu.edu.hk[2], andreas.bulling@vis.uni-stuttgart.de[3]

---

**Abstract**

Internal thought refers to the process of directing attention away from a primary visual task to internal cognitive processing. Internal thought is a pervasive mental activity and closely related to primary task performance. As such, automatic detection of internal thought has significant potential for user modelling in intelligent interfaces, particularly for e-learning applications. Despite the close link between the eyes and the human mind, only few studies have investigated vergence behaviour during internal thought and none has studied moment-to-moment detection of internal thought from gaze. While prior studies relied on long-term data analysis and required a large number of gaze characteristics, we describe a novel method that is computationally light-weight and that only requires eye vergence information that is readily available from binocular eye trackers. We further propose a novel paradigm to obtain ground truth internal thought annotations that exploits human blur perception. We evaluate our method for three increasingly challenging detection tasks: (1) during a controlled math-solving task, (2) during natural viewing of lecture videos, and (3) during daily activities, such as coding, browsing, and reading. Results from these evaluations demonstrate the performance and robustness of vergence-based detection of internal thought and, as such, open up new directions for research on interfaces that adapt to shifts of mental attention.

---

## 1. Introduction

While a large number of works have studied attention shifts between external stimuli, shifts from external stimuli to internal thought has only been studied rather recently (Smallwood & Schooler, 2015). Detecting the attention shift in a learning environment is of particular importance. The *attention shift* in this paper refers to the *shift of attentional focus away from the visual task towards learners' internal thoughts*. Specifically, to be in line with the general online learning scenarios, we define the *visual task* as attending to the visual learning material and catching every conveyed message. In practice, learners' *internal thoughts* can be task-related, e.g. goal-directed thought, as well as task-unrelated, i.e. *mind wandering* (Smallwood, O'Connor, Sudberry, Haskell, & Ballantyne, 2004; Schooler et al., 2011; Christoff, Irving, Fox, Spreng, & Andrews-Hanna, 2016). Although a number of internal thoughts can be positive to learning, it is also highly important for an intelligent system to be aware of learners' attention shift, so as to allow them to review the important missing information or to build the learners' attentional profiles. Moreover, mind wandering, a common form of internal thoughts, was showed to be pervasive (over 50% of the time) during everyday activities (Killingsworth & Gilbert, 2010) and it oftentimes decreases learning performance (Olney, Risko, D'Mello, & Graesser, 2015). As such, moment-to-moment detection of internal thought is an essential but currently missing technique towards understanding and improving the experiences of learning and teaching, for example via early intervention and attention (re)direction (D'Mello, 2016), task rearrangement (Dingler, 2016) or adapting learning material (Cheng, Sun, Sun, Yee, & Dey, 2015; Kardan & Conati, 2015).

Given the importance of attention understanding for computer-based learning as well as the rapid emergence of e-learning, previous work has explored attention detection based on gaze patterns and pupil dilation. However, the vast majority of these works focused on reading (Franklin, Broadway, Mrazek, Smallwood, & Schooler, 2013; Li, Ngai, Leong, & Chan, 2016; Bixler & D'Mello, 2016; Faber, Bixler, & D'Mello, 2017) for which gaze patterns are rather unique. In contrast, far fewer works have aimed to detect attention in other contexts, such as movie watching (Mills, Bixler, Wang, & Mello, 2016) or during interactions with a tutoring system (Hutt, Mills, White, Donnelly, & D'Mello, 2016), and no effort has been made to achieve moment-to-moment detection.

We propose two major improvements over the state of the art. First, inspired by recent psychological studies that observed vergence changes (i.e. both eyes either rotating inwards or outwards) during internally directed cognition (Benedek, Stoiser, Walcher, & Körner, 2017; Walcher, Körner, & Benedek, 2017), we propose the detection of internal thought from eye vergence behaviour. This is also supported by the phenomenon of "staring into space" (Walcher et al., 2017) and the fact that people in a visually relaxed state may transit their visual focus to a resting state called *tonic vergence* (Toates, 1974). Unlike previous methods that relied on eye movement analysis over a long period of time, such as the number of fixations (Bixler & D'Mello, 2016)(Hutt et al., 2016) or erratic fixational patterns (Reichle, Reineberg, & Schooler, 2010), the observation of vergence allows for moment-to-moment detection of internal thought from a one-second sliding time window. It may therefore explain temporal and causal changes of human performance (Olney et al., 2015), search intents (Sattar, Muller, Fritz, & Bulling, 2015; Sattar, Bulling, & Fritz, 2017), and attention (Smallwood & Schooler, 2006; Xiao & Wang, 2017) in various activities. To the best of our knowledge, this is the first work to leverage eye vergence behaviour for moment-to-moment detection of internal thought.

Second, this paper proposes a novel experimental paradigm for attention studies that provides fine-grained annotation of internal thought (see Fig 1). Although post-hoc self-reports provide an undoubtedly useful and valid measurement of attentional state (Smallwood, Davies, et al., 2004; Smallwood & Schooler, 2006), they are generally associated with an entire time segment, thus lacking information on the precise start and end of internal thought. Existing methods also rely on participants' awareness of their state of mind and perception of time, which can be subjective and error-prone (Smallwood & Schooler, 2006). In contrast, we exploit human perception of a *gradual blurring effect* to estimate the start and precisely measure the end of internal thought. This method can generate reliable and fine-grained annotations for internal thought and opens new avenues for research in the computer-based education, human-computer interaction, and neuroscience communities.

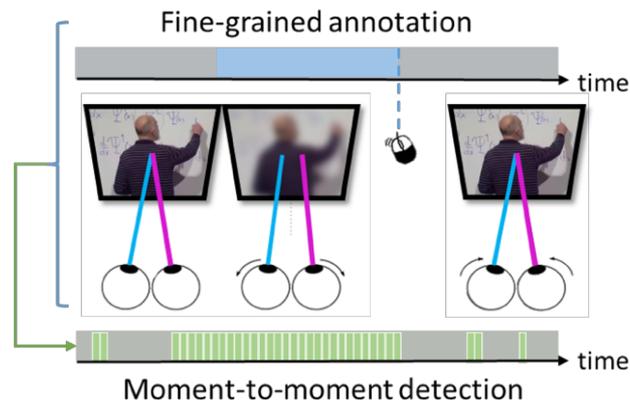

Fig 1. We propose a novel annotation paradigm for fine-grained internal thought annotation and a novel method that uses eye vergence behaviour for moment-to-moment detection of internal thought.

The specific contributions of our work are three-fold. First, we propose a novel experimental paradigm for studies of internal thought without intrusive intervention. Second, we propose and explore the use of eye vergence behaviour for light-weight and moment-to-moment detection of internal thought. Third, we conduct in-depth experimental evaluations to (1) validate our experimental paradigm, (2) evaluate our method for spontaneous internal thought detection during lecture viewing, and (3) in an in-situ environment across daily activities.

## 2. Related Work

We start with the discussion of prior works on 1) gaze-based detection of mental states with a particular focus on 2) attention detection, as well as 3) exploring the link between eye vergence and attention.

## 2.1. Gaze-Based Detection of Mental States

Numerous works have investigated the detection of attentive and mental states from eye characteristics and gaze behaviour, including pupil dilation (Franklin et al., 2013; Pfleging, Fekety, Schmidt, & Kun, 2016) (Unsworth & Robison, 2016; Toker & Conati, 2017), blinks (Oh, Jeong, & Jeong, 2012), fixations (Smilek, Carriere, & Cheyne, 2010; Foulsham, Farley, & Kingstone, 2013; Frank, Nara, Zavagnin, Touron, & Kane, 2015), scanpaths (Goldberg & Kotval, 1999; Mishra, Kanojia, Nagar, Dey, & Bhattacharyya, 2017), and saccades (Huang, Li, Ngai, & Leong, 2016; Li et al., 2016). While pupil diameter is closely related to cognitive states, it is sensitive to ambient illumination and requires expensive equipment for measurement (Bixler & D'Mello, 2016). Huang, Li, et al. (2016) studied the fixations and saccadic movements near the moment of a mouse-click for stress recognition. Li et al. (2016) investigated the recognition of reading comprehension based on fixations, saccades, and blinks. Besides these gaze features, Bulling and Roggen (2011) also included wordbooks of saccade sequences to study visual memory recall processes. Bulling, Ward, and Gellersen (2012) studied eye and head movement for reading activities recognition. Hoppe, Loetscher, Morey, and Bulling (2015) applied similar technique for curiosity recognition from eye movement. Steichen, Wu, Toker, Conati, and Carenini (2014) leveraged differential sequence mining of eye gaze data to identify user and task patterns. Mishra et al. (2017) measured reading effort by calculating scanpath complexity. Goldberg and Kotval (1999) found scanpaths and the fixation/saccade ratio are associated with cognitive behaviour underlying visual search.

Despite these successes, prior findings on the link between eye characteristics and mental states are not entirely consistent. For example, Oh et al. (2012) found that eye blinks were suppressed as the difficulty of an auditory tone-counting task increased while Andrzejewska and Stolińska (2013) showed that task difficulty was not correlated with eye blinks. Similarly, while Frank et al. (2015) showed that mind wandering leads to longer fixation durations, Foulsham et al. (2013) found that it could also result in more fixations and Smilek et al. (2010) suggested that eyes fixate less often and shorter during mind wandering. In summary, while conventional eye characteristics seem to somehow correlate with mental states, they can be task-specific and only offer limited generalizability.

## 2.2. Gaze-Based Attention Detection

The detection of learners' attention or inattention is essential to intelligent e-learning systems. Conati, Aleven, and Mitrovic (2013) presented an extensive review on the studies of using gaze data to analyse, model and respond to learners' attentional states. In practice, learners' inattention can either be overt or covert. Overt (in)attention denotes learners intentionally direct attention to or away from the task. It is relatively straightforward to detect given the eye-tracking data. For instance, D'Mello, Olney, Williams, and Hays (2012) measured the degree of disengagement by the eye looking away from the tutoring system. Navalpakkam, Kumar, Li, and Sivakumar (2012) analysed visual attention to identify the visual and semantic importance in the multi-item interface. Another important line of work was attentive user interfaces proposed by (Vertegaal, 2003; Bulling, 2016). In contrast, detecting covert inattention, i.e. mind wandering, is more challenging but interesting to us.

Most previous works focused on understanding gaze behaviour during mind wandering and only a few explored the inverse, i.e. gaze-based mind wandering detection. Franklin, Smallwood, and Schooler (2011) made the first attempt in a constrained word-by-word reading paradigm. There was a line of works investigated mind wandering detection based on the statistics of fixations, saccades, blinks, and pupil dilation. Bixler and D'Mello (2016) compared content-dependent and -independent gaze features during natural reading. Faber et al. (2017) recently suggested that using content-independent gaze features with a 12s window can perform well during reading. Mills et al. (2016) extended the study to film watching. Hutt et al. (2016) investigated interactions with a tutoring system that presented biology topics with animation, and later on lecture viewing (Hutt et al., 2017). They also found that content-independent gaze features performed consistently better and showed that longer windows (20-30s) were preferable for non-reading tasks.

These studies yielded two key insights. First, small windows (≤10s) may contain insufficient fixation and saccade information for covert inattention detection (Bixler & D'Mello, 2016; Hutt et al., 2016). With two exceptions (Bixler & D'Mello, 2016; Mills et al., 2016), recognition accuracy dropped with a decrease in window size (Hutt et al., 2016; Faber et al., 2017; Hutt et al., 2017). This limitation constrains the moment-to-moment detection of internal thought and further motivates the use of eye vergence features. Second, content-dependent features contribute little to the classification accuracy across reading, film, and lecture viewing (Bixler & D'Mello, 2016; Mills et al., 2016Hutt et al., 2017). Therefore, in this paper, we also focus on content-independent features.

### 2.3. Eye Vergence and Attention

Prior works have studied eye vergence and attention. Solé Puig, Pérez Zapata, Aznar-Casanova, and Supèr (2013) suggested that eye vergence is linked with covert spatial attention, where human visual attention is directed to other visual stimuli in peripheral vision without overt eye movement of orienting. Please note that the focus of their study was *visuospatial attention* (i.e. the spatial shift of visual focus), rather than the change of cognitive focus in the current study. Lately, their further studies also showed that eye vergence behaviour reflecting covert spatial attention can be used to identify children with attention deficit hyperactivity disorder (Solé Puig et al., 2015; Varela Casal et al., 2018), whose symptoms include inattention, hyperactivity, and impulsivity.

In contrast to the studies on covert spatial attention, two recent psychological studies explored the link between eye vergence and internal cognition, which well supports our idea of vergence-based internal thought detection. Walcher et al. (2017) found that internally directed cognition leads to increased variability in eye vergence in a letter-by-letter reading task. Benedek et al. (2017) showed a higher variability and smaller eye vergence during internally directed cognition in anagram and sentence generation tasks. However, unlike this study, their objective was to investigate behavioural patterns during *deliberate* (rather than *spontaneous*) internal cognition, and they did not perform automatic detection. Further, their methods also required extended observation of gaze behaviour over 20-30s.

## 3. Automatic Detection of Internal Thought

According to the attentional framework described by D'Mello (2016), attention to visual learning material can take three major forms (see also Fig 2): (1) *focused attention*: learners completely focus on one part of the visual task, e.g. the lecturer; (2) *alternating attention*: learners switch attention between different parts of the visual material, e.g. between the lecturer and the notes on board; and (3) *divided attention*: learners attend to visual information and meanwhile process the narration. When attention shifts away from the visual task, it can be either *overtly* distracted or *covertly* shifted to internal thoughts. In this work we aim to identify shifts of attention from the visual task (i.e. external stimuli) to internal thought.

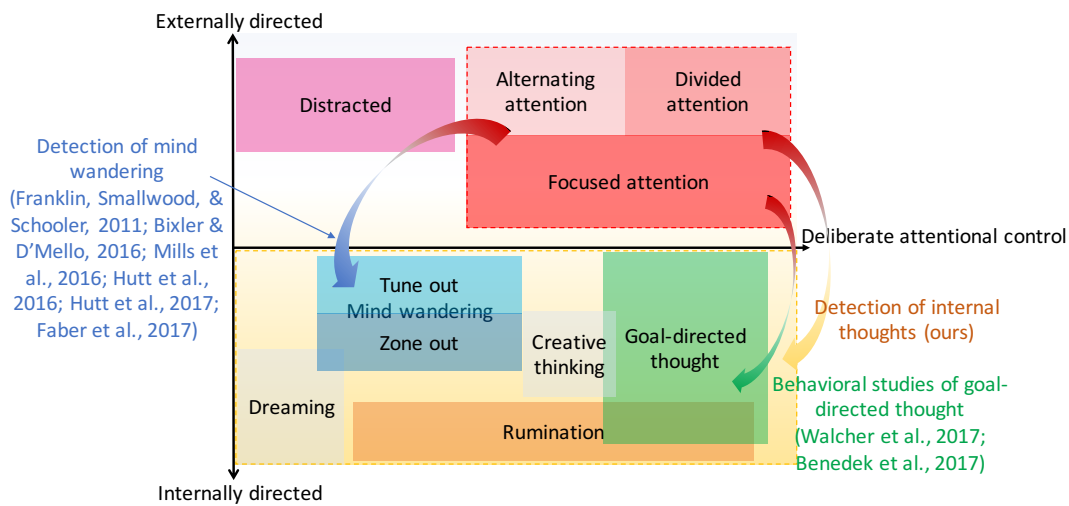

Fig 2. Conceptual space of learners' attention to the visual task in a learning environment. We aim to identify learners' attention shift from the visual task (i.e. external stimuli) to internal thought. We also highlight some of the gaze-based studies in the figure.

Christoff, Irving, Fox, Spreng, and Andrews-Hanna (2016) proposed a conceptual space of internal thoughts that includes goal-directed thought, creative thinking, mind wandering, dreaming and rumination. The most well-studied topic is the detection of mind wandering, generally defined as task-unrelated thought (Smallwood & Schooler, 2015) or perceptually decoupled thought (Schooler et al., 2011). However, knowing when learners have dived into a task-related thought and stopped receiving external messages is similarly important for an intelligent tutoring system to provide effective interventions. Therefore, our goal is to detect attention shift from visual learning material to different internal thoughts, including goal-directed thought and mind wandering.

There are two points worth noting: (1) externally and internally directed thoughts often co-occur (Smallwood & Schooler, 2006; Dixon, Fox, & Christoff, 2014). As such, we aim to only detect the period when the focus of attention is shifted from the visual task to internal thought. (2) Attention shift in this paper refers to the shift from external visual task to internal thought. It is different from the *covert attention* shift (Engbert & Kliegl, 2003; Solé Puig et al., 2013), which refers to a spatial shift of visual attention. Previous studies have relied on long-term observations of various eye movement characteristics to infer learners' attention (Hutt et al., 2016; Faber et al., 2017; Hutt et al., 2017), however, such characteristics often have inextricable links with multiple factors, including human intention and interface layout. In contrast, we believe eye vergence behaviour can be more indicative of attention and robust to biases. Therefore, we exploit eye vergence for moment-to-moment detection of internal thought. Vergence Behaviour in Eye-Tracking Data

Vergence refers to eye movements where the eyes move in opposite directions (Toates, 1974). Focusing on a close object makes the eyes rotate towards each other (converge), while focusing on a more distant object will make the eyes rotate away from each other (diverge). As a first step towards vergence-based detection of internal thought, we recorded pilot gaze data during daily computer activities using a remote eye tracker mounted below a monitor. When visualizing the on-screen gaze estimates for the right and left eye, we observed that vergence has several distinct attributes, such as disparity, direction, and dispersion. Fig 3 shows sample gaze data of the left (cyan) and right (purple) eye within a 1s window for (a-d) increasing disparity, (c) and (d) torsional vergence, (e-h) horizontal and vertical vergence, and (i-l) increasing degrees of dispersion.

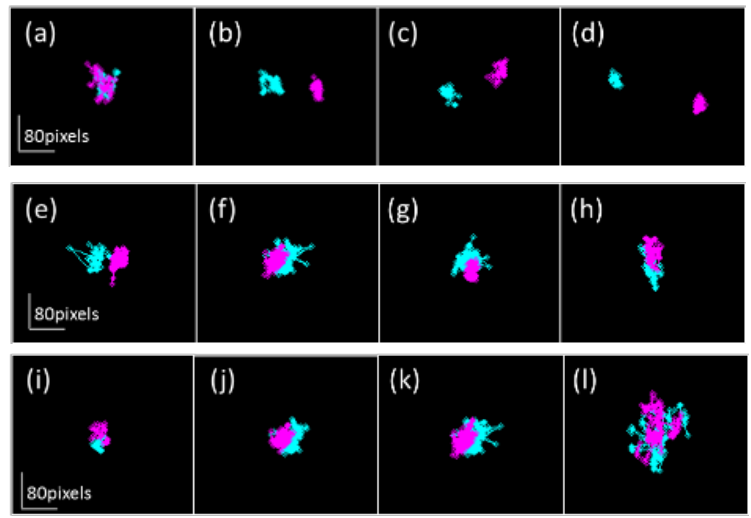

Fig 3. On-screen gaze estimation samples of the left (cyan) and right (purple) eyes within a 1s window for (a-d) increasing disparity, (c-d) torsional vergence, (e-h) horizontal and vertical vergence, and (i-l) increasing degrees of dispersion. The orthogonal lines indicate a scale of 80 pixels.

## 3.1. Eye Vergence Feature Extraction

Based on these observations, we develop two groups of vergence features: The first is extracted from pairs of left and right eye gaze estimates, while the second is extracted from fixations. The pair-based vergence features are extracted by identifying all valid gaze point pairs in the time window, and then calculating the mean and standard deviation (SD) of their distances and angles.

To calculate the fixation-based vergence features, we first resample the eye tracking data at a fixed frame rate (60FPS) and filter the data (Casiez, Roussel, & Vogel, 2012). We then detect fixations using the dispersion-threshold identification (I-DT) algorithm (Salvucci & Goldberg, 2000). We use a duration threshold of 80ms (Hansen & Ji, 2010) and a dispersion threshold of 80 pixels (≈1° in our experimental setting). We measure the disparity and angle between the centroids of the fixation points for the left and right eye. Similarly, we extract the minimal bounding circles for the corresponding gaze points and calculate their centre distance and angle. Additionally, we included the centre distance normalized by the sum of the radii of the bounding circles.

Although the disparity in screen coordinates between the gaze estimates of the left and right eyes provides a direct representation of the vergence in degrees, the distance to the screen can delineate the visual focus more precisely. We use a simplified model (see Fig 4) extended from Kudo et al.'s work (Kudo et al., 2013) to approximate the visual focus displacement from the screen as

$$d = \begin{cases} E \cdot D / (PD - E), & \text{divergence} \\ -E \cdot D / (PD + E), & \text{convergence} \end{cases}$$

where $D$ is the eye-to-screen distance, $PD$ denotes the pupillary distance and $E = \beta \|\boldsymbol{g}^L - \boldsymbol{g}^R\|_2$ indicates the gaze disparity in the world coordinate, which is transformed from the gaze disparity $\|\boldsymbol{g}^L - \boldsymbol{g}^R\|_2$ in the screen coordinate to the world coordinate by the a constant $\beta$(=0.283). The condition of divergence or convergence is determined by the horizontal locations of the left and right gaze estimates.

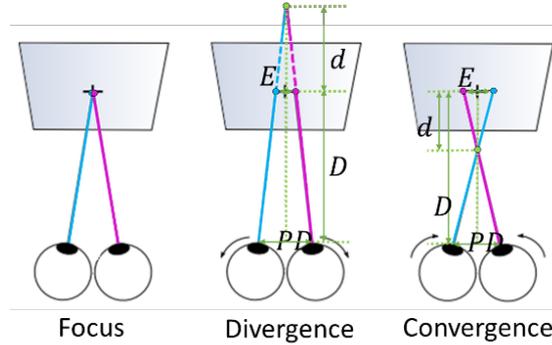

Fig 4. A simplified model for displacement estimation between visual focus and screen surface.

Table 1. Gaze features used in our evaluations. Shaded cells are features described in prior work (Bixler & D'Mello, 2016; Mills et al., 2016; Hutt et al., 2016; Hutt et al., 2017; Faber et al., 2017).

| Type | Feature description |
|---|---|
| Vergence and distance (17) | Disparity of gaze point pairs (mean, SD); gaze focus distance from screen (mean, SD) |
| | Disparities of gaze point sets – centroid distance; centre distance of minimal bounding circles; and centre distance over sum of radii |
| | Direction/angle of gaze point pairs (mean, SD); and their mean centroid and centre angles |
| | Distance between eyes and screen; pupillary distance (mean, SD) |
| Fixation (13) | Radii of minimal bounding circles |
| | **Duration of fixation**; fixation duration and total number over the window |
| | Duration ratio of fixation over saccade |
| Saccade (86) | **Duration, length, and velocity of Saccades**; duration and total number over the window |
| | **Angles in degrees between saccades relative to the x-axis and to the previous saccade** |
| | Proportion of horizontal saccades with angles of [-30°, 30°] relative to the x-axis |
| Blink (4) | Duration of blink (mean, SD); blink duration and total number over the window |

The bold features use multiple descriptive statistics, including mean, standard deviation (SD), median, min, max, range, kurtosis, and skewness. Numbers in brackets indicate the number of features.

Table 1 summarizes all features used in this work, including the proposed vergence features and the features proposed in prior work. In total, we extracted 120 features. The shaded features (classic feature set) are adapted from previous works and mainly cover fixations, saccades, and blinks (Bixler & D'Mello, 2016; Mills et al., 2016; Hutt et al., 2016; Hutt et al., 2017; Faber et al., 2017). Following their methods, the bold attributes (i.e. fixation duration, saccade duration, length and velocity, angle between saccades) use multiple descriptive statistics as features, including mean, standard deviation, median, min, max, range, kurtosis, and skewness. Slightly different from studies that were only interested in the overall movement of both eyes, we calculate the saccadic features (duration, length, velocity, and angles) separately for each eye to capture the different oculomotor behaviours between left and right eye. In addition, as our goal is moment-to-moment detection with a window of less than 1s, most of our new features use only mean and SD. Finally, since we aim for a user-independent model that needs to generalize to completely unknown users, within-participant feature normalization is not performed (i.e. raw measurements are used).

We designed a series of user studies to evaluate the proposed vergence-based method for detecting internal thought. In the following, we first evaluate our method for detecting goal-directed internal thought in a controlled math-solving task. For an in-depth analysis of the realistic and general behaviours, we then present our study on the proposed experimental paradigm for non-intrusive annotation of internal thought, followed by the evaluation of its use during natural viewing of lecture videos. Finally, we discuss about a further investigation of vergence-based detection of internal thought across different activities, including browsing, coding, and reading.

**Study I: Detecting Internal Thought during a Controlled Math-Solving Task**

As a first step towards the detection of internal thought, we evaluated our vergence-based detection method by inducing the attention shift from a visual task to the goal-directed thought through an additional mental math-solving task. This study fills a critical gap in prior work by investigating, for the first time, the detection of goal-directed thought (see Fig 2).

*3.2. Experimental Procedure*

The study consisted of two sessions for each participant. In both sessions, similar to (Varela Casal et al., 2018), participants were instructed to fixate on an image target. Specifically, we used an image of a face, since it is one of the most often-encountered images in multimedia and it is well known to be a salient object that naturally attracts attention (Xiaohui Shen & Ying Wu, 2012). Compared to the traditional covert spatial cueing experiment with small fixation targets (Engbert & Kliegl, 2003; Solé Puig et al., 2013), this setting allowed for natural fixational behaviours closer to those in real-world situations, such as lecture viewing. We used a baby's face to reduce potential visual biases caused by a particular gender. To account for the possible impact from the location and size of the visual target, we displayed the target in five locations (four corners and the centre) on the screen, in two sizes (20x20 and 100x100 pixels), twice, for 10s each time, for each session. This yielded 200s of data per participant and session.

Participants performed both sessions in random order. In one of the sessions, we used mental arithmetic to induce goal-directed thought in the participant, as commonly used as a stimulus in human cognitive processing studies (Huang, Li, et al., 2016). Specifically, while fixating on the visual target, participants were asked to calculate and speak aloud the answers to a series of arithmetic questions given verbally by the experimenter. The verbal interaction with the experimenter directed participants away from the visual task. To reduce possible biases from mental stress and frustration, we constrained the difficulty of the mental arithmetic by using 2-digit additions and subtractions. For the same reason, we did not set a time limit and did not provide feedback on response accuracy. However, a question was asked right after the participant responded, so as to ensure that the math-solving task occupied the major cognitive capacity of the participants and directed their attention away from the visual target.

We recruited 16 participants for this study (10 female; mean age: 27.0, SD: 4.1). Participants have seated about 60 cm away from the monitor in a comfortable posture without using a chin rest. We used a 22" monitor at $1680 \times 1050$ resolution to display the stimuli in full-screen mode and a Tobii EyeX remote eye tracker recording binocular gaze at around 60 FPS. The eye tracker was carefully calibrated before each session to ensure the binocular tracking performance. The accuracy of this eye tracker is around 0.6° and 0.9° in the x- and y-direction, respectively, and its precisions in both direction are around 0.9° (Feit et al., 2017). A post-experiment interview confirmed that participants were able to fully focus on the visual target in the non-math sections, and that the mental arithmetic required much cognitive effort. We therefore annotated the data in the mental arithmetic sessions as attention shifts to "internal thought", and the others as "visually on-task". This resulted in an evenly distributed dataset.

### 3.3. Results and Discussion

To evaluate the proposed method for moment-to-moment detection of internal thought, we used sliding windows of four sizes (250, 500, 750, and 1000ms; step sizes of 1/4th of the respective window size) to generate instances for the classification between internal thought and visually on-task. We used the Random Forest classifier (Breiman & Leo Breiman, 2001) and performed a leave-one-participant-out cross-validation in which the data from $n$-1 participants was used for training and the remaining data for testing. This was repeated for all participants and the individual performance results were averaged. We additionally tuned the max depth of the trees through 5-fold cross-validation in the training set, with the number of trees set to 100, and the number of features in each tree int($\log_2 m + 1$), where $m$ is the number of features.

Given the similar nature of task, we compared our method to three state-of-the-art methods for mind wandering detection during reading (Bixler & D'Mello, 2016; Faber et al., 2017), lecture viewing (Hutt et al., 2017), and interacting with a tutorial system (Hutt et al., 2016). These methods use the shaded features (Rows 3-5) in Table 1, though previous work did not use blink features in non-reading tasks (Hutt et al., 2016; Hutt et al., 2017). The following results are labelled by the classifier used to generate them: Naïve Bayes (Bixler & D'Mello, 2016), Logistic (Faber et al., 2017), and Bayes Network (Hutt et al., 2016; Hutt et al., 2017). As a baseline, we also included the ZeroR method, which always predicts the major class in the training set.

Fig 5 shows the overall F1 score of the Random Forest (RF) classifier using the full feature set (red solid line) and using only our proposed 17 vergence features in Table 1 (red dashed line). We also provide the performance of the random forest classifier using only the "classic" fixation, saccade and blink features from previous work (Bixler & D'Mello, 2016; Faber et al., 2017), which is indicated by the red dotted line (RF (classic set)). Performance of the state-of-the-art methods is shown with the yellow, green and blue lines.

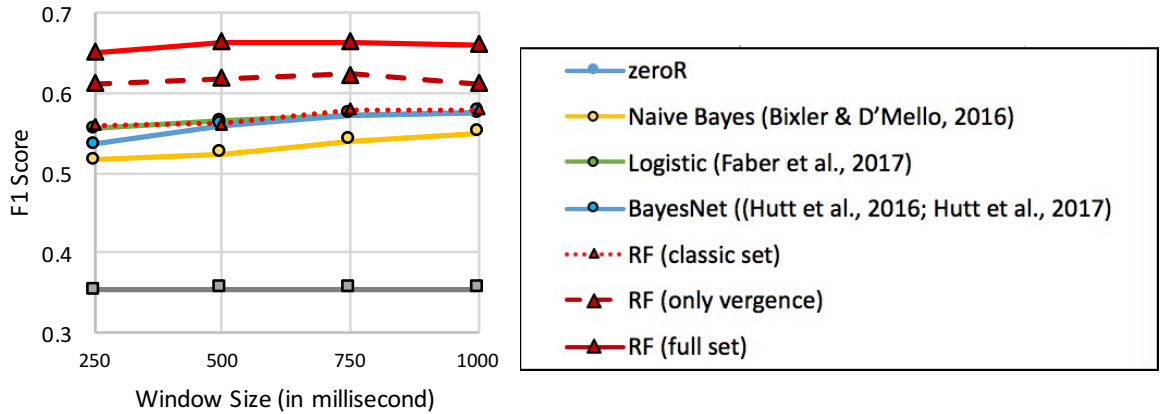

Fig 5. Weighted average F1 scores of detection of goal-directed thought. The x-axis indicates the window size. Vergence features are effective for this task, both when used alone and together with other gaze behaviours.

As can be seen from the figure, using the full feature set results in a significantly (p<0.001) higher performance than when using only the features from previous work. The state-of-the-art methods perform similarly with average F1 scores of 0.53 (Bixler & D'Mello, 2016), 0.57 (Faber et al., 2017) and 0.56 (Hutt et al., 2016; Hutt et al., 2017). With the same feature set, random forest yields a similar score (F1: 0.57). However, when given additional vergence features, it achieves a 16% increase, reaching an overall F1 of 0.66. This demonstrates the effectiveness of the vergence features. Interestingly, using only vergence features also achieves a significantly (p=0.02) higher F1 (0.62) than state of the arts.

The results also indicate that, as the window size increases, the performance achieved by the classifiers that do not use the vergence features increase slightly, though the change is not statistically significant. The performances of RF (full set) and RF (only vergence), however, are quite robust to the window size. This suggests that vergence features can capture the momentary gaze behaviour, which makes it suitable for moment-to-moment detection of internal thought. Despite this promising result, we still need to ascertain whether the gaze behaviour picked up by our model reflects the attention shift to internal thought or the difference between the single- and dual-task conditions. Thus, we perform further evaluations in more natural scenarios.

*3.4. Summary*

Vergence features contribute to improving the detection accuracy of goal-directed internal thought and they are more robust to the size of time window compared with fixations, saccades, and blinks features.

## 4. Fine-Grained Annotation of Internal Thought

Ground truth of internal thought is challenging to obtain. A related line of work is mind wandering annotation. Existing experimental paradigms are either probe-caught (Reichle et al., 2010; Bixler & D'Mello, 2016; Hutt et al., 2016; Hutt et al., 2017; Unsworth & Robison, 2016) or self-caught (Smallwood & Schooler, 2006; Reichle et al., 2010; Mills et al., 2016; Faber et al., 2017). *Probe-caught* paradigms interrupt participants intermittently during or upon the completion of a task to acquire their experience. However, it can be ambiguous whether a mind wandering episode that ends shortly before the probe should be counted or not; participants' sense of time can also affect the self-report reliability. Besides, it is almost impossible for existing paradigms to mark the start and end of mind wandering. For example, even if the participant is accurately aware of his/her own mind wandering, existing paradigms can only make a coarse-grained annotation as to whether mind wandering occurred during the time window.

In contrast, *self-caught* paradigms require participants to consciously reflect upon their experience. This alleviates the annotation precision problem, as it gives an approximation of the end. However, while mind wandering, people react relatively slowly (Unsworth & Robison, 2016; Mijović et al., 2017), and there is a high delay uncertainty between the report time and the end of the mind wandering episode. Moreover, people often fail to notice that they are mind wandering at all (Smallwood & Schooler, 2006). Likewise, probe-caught and self-caught annotations for internal thought suffer from the same problems.

*4.1. Annotation Paradigm Overview*

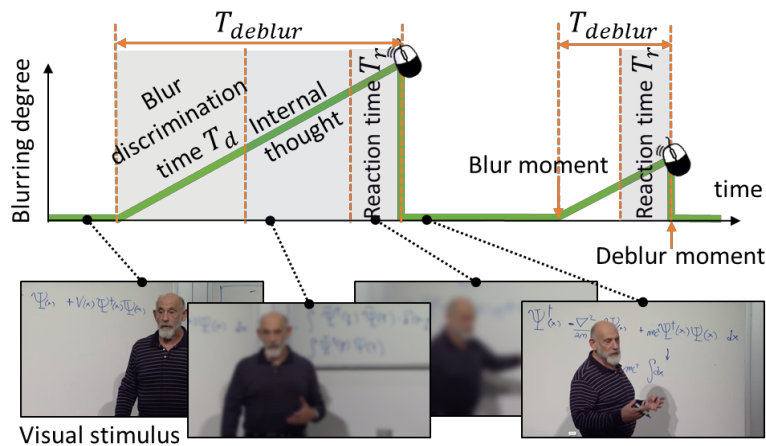

Fig 6. Paradigm to obtain fine-grained annotations of internal thought. Participants watched a video which blurs gradually at random intervals. Participants were instructed to click and deblur the video when they noticed that it was blurry. Being visually on-task may lead to a quick deblur. In contrast, people who were attending to internal thought may fail to notice the gradual blur effect and produced a slow deblur. Excluding the time needed for visual discrimination and reaction gives a conservative duration for internal thought.

To address some of these limitations, we propose a new experimental paradigm for fine-grained internal thought annotation. This paradigm (1) allows us to assess internal thought behaviour in a more fine-grained manner and (2) yields clean data for training internal thought detection methods. The key idea of our paradigm is to blur the visual stimulus (e.g. a video, document or webpage) gradually and periodically at random intervals of 10-20s (see Fig 6). Participants are instructed to click or press a key once they notice the blur effect, which immediately deblurs the stimulus. We specifically opted for gradual blurring because a sudden blurring can be visually salient and disrupt an ongoing attention shift. We used a Gaussian kernel to generate the blur effect with an aperture size of 15 pixels and with the Gaussian standard deviation, $sigma$, defined by a linear function of the time in seconds, $t$, elapsed from the blur beginning, i.e. $sigma = \alpha \cdot t$, where $\alpha$ is a factor of blurring speed.

### 4.2. Identification of the Start and End of Internal Thought

To identify the start and end of internal thought, we aimed to investigate two hypotheses:

*H1: People who are visually on-task can perceive the gradual blur effect consistently when it reaches a certain degree (i.e. $sigma = \alpha \cdot T_d$), which corresponds to the discrimination time $T_d$.*

*H2: When people's attention shifts from the visual task to internal thought, they may fail to notice the blur effect. Therefore, their time to deblur, $T_{deblur}$, is slower and greater than $T_d$.*

The key idea is that there is a discrimination time, $T_d$, by which the blurring effect should be so obvious that people who are visually on-task can notice it immediately, and thus click the key to deblur the visual stimulus. Since deblur action only happens if the participant is visually on-task, it marks the end of internal thought. While $T_{deblur} \leq T_d$ may be mainly due to the inability of the eye to discriminate small amounts of blur, $T_{deblur} > T_d$ may imply the attention shift to internal thought. In other words, $T_d$ provides a conservative start of internal thought. Participants' reaction time is also factored into this paradigm. Based on the human reaction speed reported in prior studies (Unsworth & Robison, 2016; Mijović et al., 2017), a small response time $T_r$=0.3s is included just before the deblur action is triggered (see Fig 6). Given this paradigm, the period of internal thought is annotated from $T_d$ after the blur effect begins, to $T_r$ before the deblur moment as shown in Fig 6. Our paradigm is therefore able to provide a conservative estimate for internal thought at a fine-grained resolution.

## 5. Study II: Understanding the Perception of Gradual Blurring

We first investigated hypothesis *H1*. We studied the dominant factor for blur perception and deblur action to determine the discrimination time $T_d$.

### 5.1. Procedure and Setting for Gradual Blur Effect Perception

Human blur perception is affected by multiple factors, including motion, luminance, depth, and screen attributes (Watson & Ahumada, 2011). In real use, screen attributes are a factor we can hardly control for. Since the depth in our video stimuli is rather consistent, this study focuses on the remaining three factors of blurring speed, scene motion, and luminance.

For simplicity, we studied deblur actions under three conditions for each factor. Specifically, we evaluated three levels of blurring speed: fast ($\alpha = 2$), medium ($\alpha = 1$), and slow ($\alpha = 0.5$). We extracted three scene motion levels from our video stimuli: small (only lecture slides were shown), medium (a small window in the corner of the screen showed the lecturer; the rest of the screen showed the slides), and large (the screen showed the lecturer, as in Fig 6). We also adjusted the pixel intensity of the videos to obtain three levels of luminance: dark (-50), normal (no change), and bright (+50).

We recruited 11 participants for this blur perception experiment (seven female; mean age: 27.2, SD: 4.4). Participants were instructed to stay focused and click to deblur the video as soon as they perceived the blur effect, which was generated randomly after 2-5s from the beginning of each video clip. In total, we prepared 22 video clips (eight of small motion, six of medium, and eight of large), each of which lasted for 10s. With the combinations of different luminance degrees and blurring speeds, this experiment took 33 min per participant. We used the same experiment setting as that of the previous section.

## 5.2. Results and Discussions for Gradual Blur Effect Perception

We plot the probability distributions of deblur actions of all the participants under different conditions of blurring speed, motion, and luminance. Fig 7 shows their conditional probability mass functions. As expected, the blurring speed highly affects the deblur action. In general, a fast blurring speed results in a quick deblur. The slow blurring speed leads to fatter tails and greater variability, while the medium and fast have much narrower shapes. The deblur probabilities of fast and medium speeds peak at 1s and 1.2s, and the clear majority of the actions were completed before 1.5s. Based on this finding, we used the medium blurring speed ($\alpha = 1$) in our following experiment because it (1) produced a consistent deblur pattern with acceptable variability across participants and (2) posed less salient visual impact than a fast blurring speed.

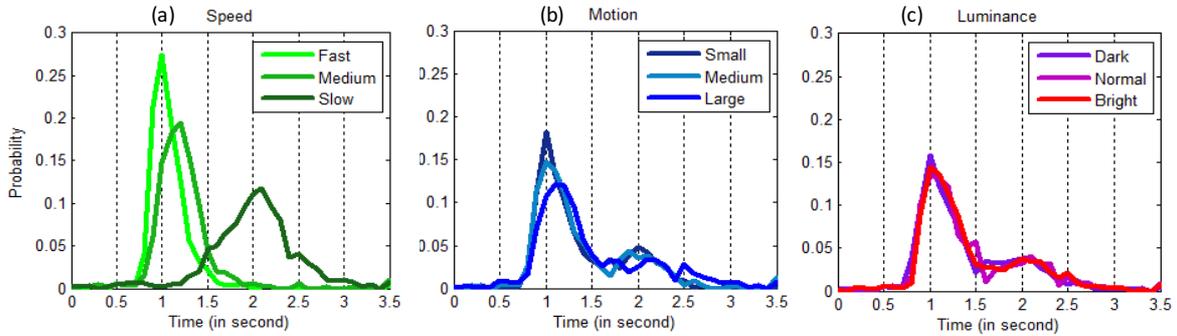

Fig 7. Probability mass functions of deblur actions under different (a) blurring speeds, (b) scene motions, and (c) luminance degrees. The scene motion and luminance do not affect the deblur action much. At a medium blurring speed, a clear majority of participants perceived the blur effect and triggered the deblur action within 1.5s.

It is interesting to see that the scene motion and luminance do not obviously affect the deblur actions. The curves in Fig 7 (b) and (c) significantly overlap. This further indicates that it is very likely that people who are visually on-task can perceive the blur effect and trigger the deblur action within 1.5s under medium blurring speed. Excluding the reaction time $T_r$ (=0.3s) gives us $T_d$=1.2s.

## 5.3. Summary

This experiment corroborates *H1*: that people visually on-task can perceive the blur effect within a consistent time. It also provides evidence to determine the discrimination time $T_d$, which facilitates the identification of the start and end of internal thought.

## 6. Study III: Detecting Internal Thought during Natural Viewing of Lecture Videos

Given the growing popularity of online courses, we evaluated our method on lecture viewing similar to Pham and Wang (2015) and Hutt et al. (2016). We first investigated hypothesis *H2* based on our experimental observations. We then present the performance of the classification of internal thought from deliberate on-task and spontaneous on-task.

## 6.1. Experimental Procedure

A common lecture timespan is around 50-60 min. Therefore, we prepared six video clips for this experiment, each of which lasted for 10 min. As discussed previously, different types of scenes do not affect blur perception much and we thus simplified the scene types. Three video clips contained only the slides view, while the other three showed the view in the lecture room. To increase the chance of capturing sufficient data of spontaneous internal thought, our videos were chosen to be challenging to follow: the topics were difficult, such as Riemann Hypothesis and Fermat's Last Theorem, and we also slowed down the video clips to 85% of the original speed.

Our post-experiment interview confirmed that participants indeed found difficulty to visually attend to these videos and their minds wandered frequently.

We recruited 24 participants (12 female; mean age: 25.6, SD: 3.1) for this experiment, and used the same experiment hardware setting as previous experiments. We recorded the screen video as well as the facial video using a webcam mounted on the monitor. Due to the decrease of eye tracking accuracy near the screen edges (Feit et al., 2017), we placed the video stimulus in the centre of the screen with a size of 18.1 cm by 10.2 cm, which is the default size of YouTube video in our setting. Participants were allowed to sit freely but recommended to avoid significant head movements. Applying our internal thought annotation paradigm produced 2943s of "internal thought" data across participants.

In addition, we recruited another 16 participants (six female; mean age: 26.9, SD:2.0) and instructed them to view shorter editions of the 6 video clips (30s each) in a complete focused attention. This produced 2880s of clean training data of *deliberate* "visually on-task" across participants. Although the deliberate data might sound artificial, it is an effective way to obtain clean and reliable visually on-task data. We will show in the later analysis that this clean data is consistent with the spontaneous data.

### 6.2. Understanding Internal Thought in Lecture Viewing

There are several interesting findings from our data. First, fixations during lecture viewing are very different from those during reading. Rayner pointed out that fixations in reading usually have a duration of approximately 200-250ms (Rayner, 1998). Fig 8 (a) shows the histogram of fixation duration across all participants throughout the lecture viewing experiment. Although the distribution peaks around 250ms, there is a long tail reaching over 3s, indicating that there are many more long fixations (over 250ms) during lecture viewing than reading. This suggests that eye vergence features can be more indicative and effective for moment-to-moment detection of internal thought than the classic gaze movement features, which require patterns of multiple fixations and saccadic movements.

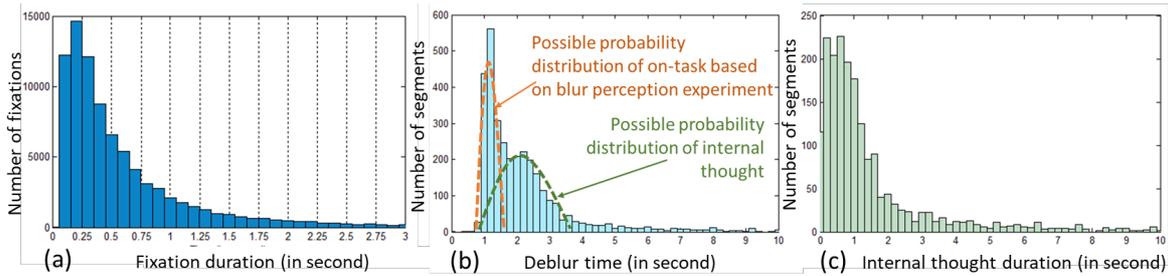

Fig 8. (a) Histogram of fixation durations across participants throughout the lecture viewing experiment. (b) Histogram of deblur time $T_{deblur}$ from the beginning of each blur effect. (c) Histogram of the conservative duration of internal thought.

The second interesting finding is that there are many slow deblur actions ($T_{deblur} > T_d = 1.5$s) in the lecture viewing data, which verifies our hypothesis *H2* that people who are mind wandering fail to notice the blur effect and delay the deblur. Fig 8 (b) presents the histogram of the deblur time across participants. We see that $T_{deblur}$ has a bimodal distribution, which is illustrated by the orange and green dashed lines. There is a clear peak near 1s. This is in accordance with the conditional probability mass functions (Fig 7) when participants are visually on-task. The other peak is at around 2.5s, which probably reflects the deblur distribution during internal thought. Since these two peaks are close to each other and suggest two highly overlapping distributions, to obtain reliable annotated data, we only assume that a long $T_{deblur}$ is indicative of internal thought but not the other way around.

Third, our data suggests that a small time window can be appropriate to detect the fragmented internal thought during lecture viewing. After identifying the start and end and subtracting $T_d$ and $T_r$ from the long $T_{deblur}$ (> $T_d$) segments, we obtain the conservative duration of internal thoughts (see Fig 8 (c)). Removing outliers of over 10s (~2% of the internal thought segments; since one participant actually fell asleep), the average internal thought duration is 1.4s (SD: 1.7s). These durations are conservatively estimated, but it still implies that internal thought during lecture viewing can be short and fragmented, around 1-3s. Although large time window (20-30s) used in prior studies achieved great success in detecting the overall attentional state (Bixler & D'Mello, 2016; Hutt et al., 2016; Hutt et al., 2017; Faber et al., 2017), a small time window has the potential of a finer-grained detection of internal thought in a moment-to-moment fashion.

### 6.3. Results and Discussion

We now present a quantitative evaluation of internal thought detection. Following the practice in Study I, we used sliding windows within 1s (250, 500, 750, and 1000ms) to generate instances for training and testing. The resultant datasets ranged between 60,000 instances (250ms) and 8,500 instances (1000ms) and contained on average 45.1% (SD: 3.5%) of internal thought and 54.9% of deliberate on-task data. We conducted a leave-one-participant-out cross-validation comparing with state-of-the-art methods (Bixler & D'Mello, 2016; Hutt et al., 2016; Hutt et al., 2017; Faber et al., 2017).

#### 6.3.1. Classification between internal thought and deliberate on-task

We first investigate whether our classifier can distinguish between internal thought and deliberate on-task. Fig 9 shows the F1 score of the Random Forest (RF) classifier against state of the art and baseline across different window sizes. Since ZeroR heavily depends on the class distribution in training set, it can be susceptible to window size, which affects the number of instances. For example, the 250ms window leads to a very low F1 (0.04) for ZeroR (see Fig 9(a)). However, it is encouraging that all other methods are more robust to the data distribution and achieves significantly higher F1 scores than the baseline (p<0.001).

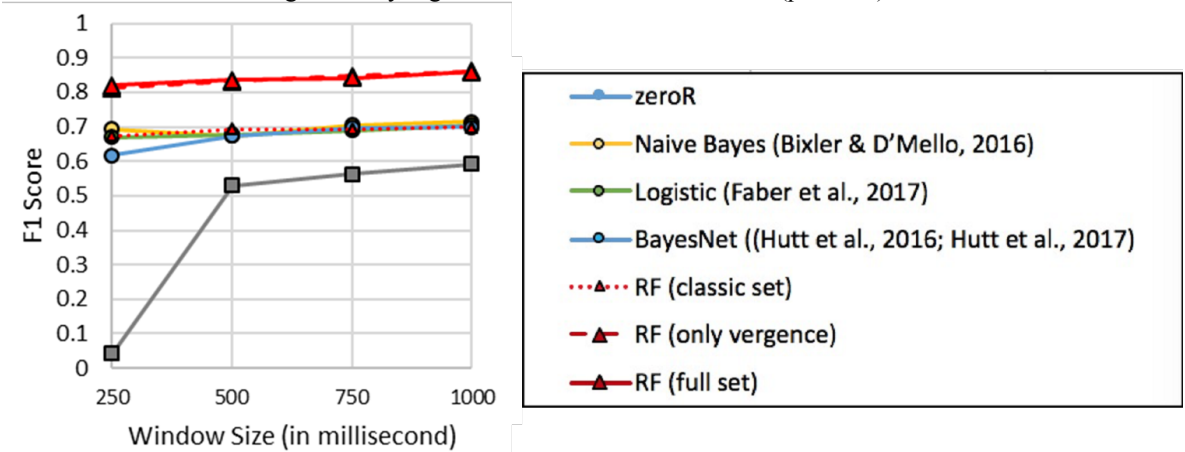

Fig 9. Weighted average F1 scores of the classification between internal thought and deliberate on-task. Using vergence only features performs comparably with that of using full feature set.

It is even more encouraging that only using vergence features performs as well as using the full feature set for RF, both reaching an average F1 score of 0.84 across different windows. This is a significantly higher performance than that achieved by the state of the art (F1 around 0.68, p<0.001). Although performance tends to increase as window size increases, the difference is not significant (p=0.09 between 250ms and 1000ms).

These results also imply that the classic features (fixations, saccades, and blinks) within 1s windows do not contribute much to the performance, at least for lecture viewing. Unlike Study I where participants were instructed to fixate on a face image, fixations, saccades and blinks in lecture viewing can be affected by numerous factors, such as video saliency and user intention. This suggests that in natural environments using vergence features for moment-to-moment detection is preferable and even more appropriate.

#### 6.3.2. Classification between internal thought and spontaneous on-task

The vergence features appear effective in distinguishing between internal thought and deliberate on-task, however, an even more interesting question is: can the deliberate on-task data contribute to the discrimination between internal thought and spontaneous on-task?

To answer this, we further propose an annotation method to obtain "spontaneous on-task" data. We postulate that participants can maintain on-task for at least 1.5s after they refocus back to the visual task. This should be a plausible assumption, since the blur perception experiment suggests that participants can sustain attention at least for a discrimination time $T_d$ (=1.5s). Therefore, we annotate *the 1.5s period starting from the end of an internal thought (i.e. $T_r$ before the end of $T_{deblur}$) as spontaneous on-task,* as illustrated in the top column in Fig 10. Extracting the features with different window sizes yields four datasets with an average 41.8% (SD:5.5%) of internal thought instances and 58.2% of spontaneous on-task instances.

However, a close scrutiny of the facial video discloses that participants were not always fully attentive after every deblur action. Although they were not completely zoned out, they could be in low arousal and low engagement, i.e. on the verge of the focused attention. Fig 10 (a,b) gives two examples of a participant's facial expressions shortly after the deblur action. It is not uncommon to see participants' eyes staying in a relaxed state and mindlessly wandering around visually salient objects. In contrast, Fig 10 (c) shows an example in a highly-engaged state with a focused vision.

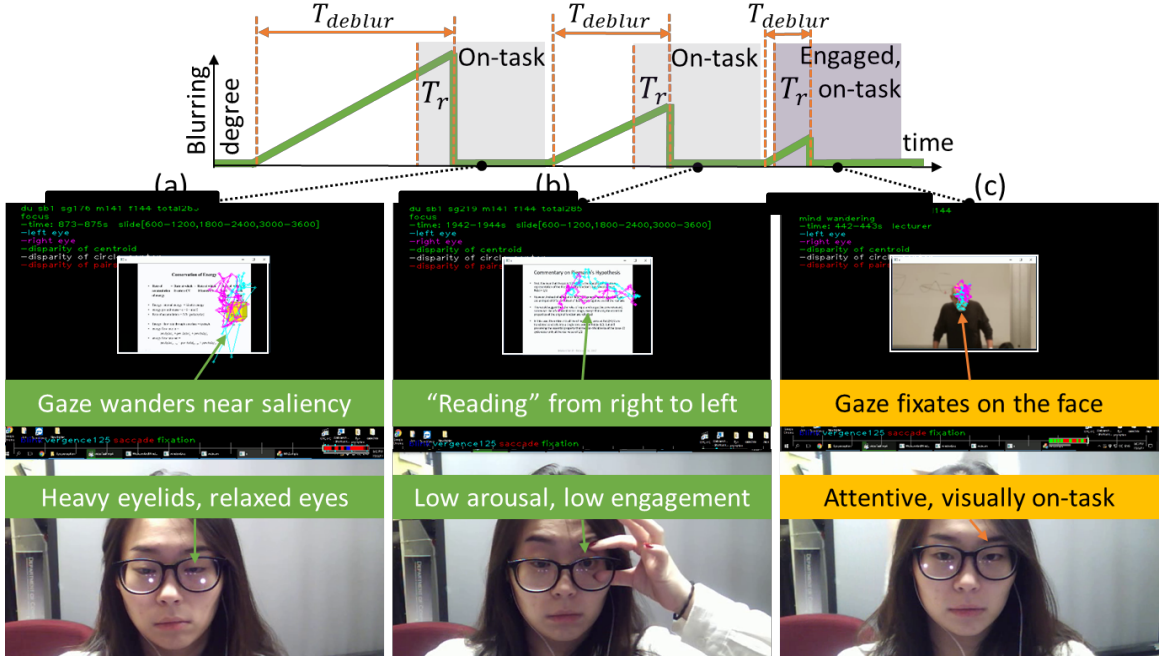

Fig 10. Illustration of spontaneous on-task annotation. The periods after long deblurs (light purple, (a) and (b)) may be in low engagement, while the one after a quick deblur (dark purple, (c)) may indicate a highly-engaged state. Left (cyan) and right (purple) eyes in low engagement might wander and diverge as during internal thought, while they were focused during the participant was attentive.

To understand the impact of this low engagement issue, we introduce an additional assumption that *a shorter $T_{deblur}$ indicates that a participant is more highly engaged and focused on the visual task.* Therefore, we extracted a fourth set of data by only including the instances where $T_{deblur} \leq 1.5$s. This gives us another four datasets with an average 62.1% (SD: 5.2%) of internal thought instances and 37.9% of spontaneous on-task instances across different windows.

Similar to the previous evaluations, we performed a leave-one-participant-out cross-validation. Specifically, on each iteration, we trained on the internal thought and spontaneous on-task data of the training participants as well as the deliberate on-task data from a separate group of participants, and tested on the left-out participant. Fig 11 (a) presents the F1 score of the RF classifier with the vergence features against the baseline and state of the arts in discriminating against spontaneous on-task and internal thought instances on the $T_{deblur} \leq 1.5$s datasets, and Fig 11 (b) on the full set with all the deblurs.

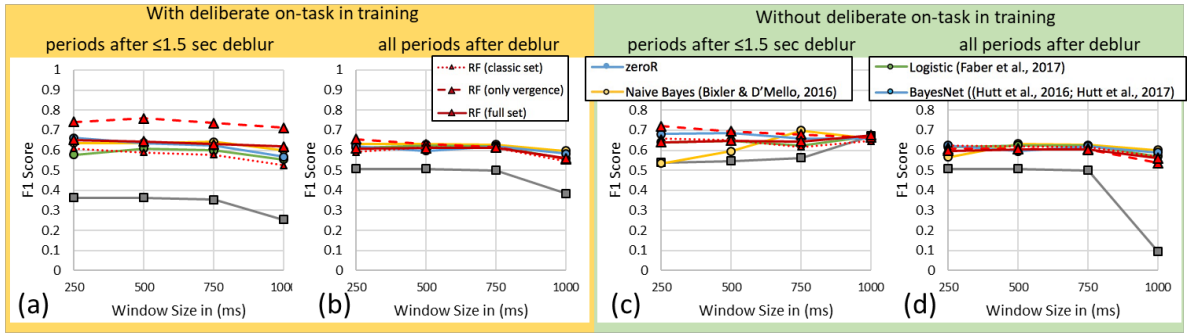

Fig 11. Weighted average F1 scores of the classification between internal thought and spontaneous on-task while training with (a-b) and without (c-d) data of deliberate on-task. Vergence features can be used to effectively distinguish internal thought from spontaneous on-task behaviour. (a) and (c) show the results of using only highly-engaged on-task data (after deblurs less than 1.5s), while (b) and (d) show the results of using on-task data annotated after all deblurs.

It is encouraging that RF (only vergence) yields an obviously higher performance than its counterparts (see Fig 11 (a)). It achieves an average F1 of 0.74, while the results of the other methods range from 0.52 to 0.66. Surprisingly, RF (full set) fails to perform as well as RF (only vergence) in this case. This is probably because the more natural contexts in this experiment made classic features more susceptible to noise. Having said that, RF (full set) with the vergence features still improve significantly over the baseline ($p \leq 0.008$). We also see that time windows do not affect the performance much. In general, this indicates that vergence features can effectively distinguish internal thought from spontaneous on-task. However, this difference decreases when tested on spontaneous on-task data extracted after all deblurs (see Fig 11 (b)). This is likely to be because some of the spontaneous on-task data from long deblur instances are likely to be on the verge of focused attention, which is challenging to discriminate for both gaze movement features and vergence features.

### 6.3.3. Contribution of training with deliberate on-task data

To evaluate the usefulness of including deliberate on-task data in training, Fig 11 (c-d) shows the performance of training without the deliberate on-task data. Although RF (only vergence) still produces the highest overall F1 (0.69) when testing on data extracted from the spontaneous on-task period after $T_{deblur} \leq 1.5s$ actions, there is only a modest performance improvement over other methods. When testing on data from the spontaneous on-task periods after all deblur actions, all the methods perform comparably, which is similar to the situation in Fig 11 (b). To conclude, the deliberate on-task data can contribute to improving the detection accuracy of internal thought and it relies on no additional annotation assumption.

### 6.4. Comparing Goal-Directed Thought in Math-Solving Task and Internal Thought in Lecture Viewing

To study the generalizability of the vergence features, we further studied internal thought detection in cross-task contexts. We evaluated the performance of training on the data of deliberate on-task and internal thought in the lecture viewing experiment (Study III), and testing on the data of on-task and goal-directed thought in the math-solving task experiment (Study I), and vice versa. A 1s window was used.

Table 2. F1 scores in the cross-tasks evaluation between math-solving task and lecture viewing experiments using a 1s window.

| Training | | State of the art | All Features | | Only Vergence | |
|---|---|---|---|---|---|---|
| | | | Math-solving | Lecture viewing | Math-solving | Lecture viewing |
| Test | Math-solving | 0.58 | 0.66 | 0.66 | 0.61 | 0.70 |
| | Lecture viewing | 0.72 | 0.64 | 0.86 | 0.60 | 0.86 |

Table 2 shows the cross-task F1 scores using both the full feature set and only vergence features. We also provide the highest within-task F1 of the state-of-the-art methods from previous works. We see that training on the lecture viewing data and testing on the math-solving task data generally out-performs within-task state of the art. Interestingly, using only vergence features achieves the best performance, and training on the lecture viewing using only vergence features produces an even higher F1 score (0.70) than that of the best within-task performance (0.66). This demonstrates the *generalizability of the vergence features*.

### 6.5. Summary

In summary, experiments in this section reveal five interesting findings. (1) The vergence features, together with Random Forest (RF (only vergence)), outperforms the state of the arts in distinguishing between internal thought and deliberate/spontaneous on-task. (2) Vergence-based features also show good generalisability across math-solving and lecture viewing tasks. (3) Training on the deliberate on-task data contributes to the recognition of spontaneous on-task, as including it in training gives a higher F1 score (0.74) than without (0.69). (4) Gaze behaviour on the verge of focused attention and internal thought can be very similar. This implies that future attention detection research can probably explore dimensional rather than categorical recognition of internal thought. (5) Shorter time windows pose only a minor impact on performance, which makes possible moment-to-moment detection of internal thought.

## 7. Study IV: Detecting Internal Thought during Daily Activities

Given the prevalence of the mind wandering (over 50% of the time) in daily activities (Killingsworth & Gilbert, 2010), we applied our vergence-based internal thought detection technique to construct a mind alert and tested it in-situ. Existing mind wandering detectors often relied on pattern analysis of multiple fixations and saccades, thus fell short of detection during lengthy fixations, in particular when users are "staring into space" (Walcher et al., 2017). To address this, we focused on the detection of internal thought during fixations. We foresee that vergence-based method can be an effective and necessary complementary solution to existing mind wandering detection methods. The detection of internal thought may sometimes lead to task-related thoughts, which are difficult to distinguish from mind wandering. To minimize the negative effect, we implemented an alert that required only a small amount of cognitive load and can be easily ignored during concentration, which is in accordance with Wickens' multiple resource theory (Wickens, 1981).

We tested the mind alert in a focused study with six participants (two female, mean age: 28.7, SD: 2.8). Each participant spent 2-6 hours performing daily activities on their own computer, including coding, browsing, and reading. A Tobii EyeX was mounted under their display and connected to a separate laptop running our mind alert. We used only vergence features extracted from a 1s sliding window. The data was sampled at 60fps; if all the frames in the window were classified as internal thought, a 0.5s audio alert was generated. The model was trained on the dataset with deliberate on-task and internal thought data during lecture viewing (see 6.3.1).

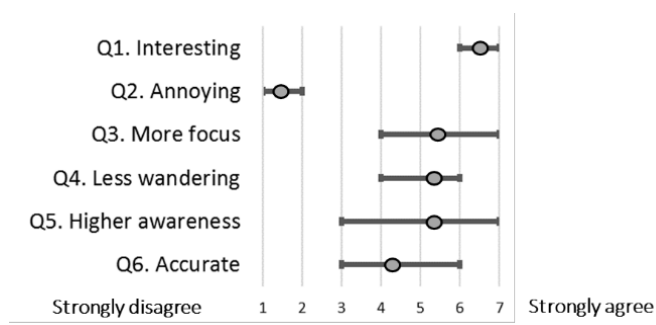

Fig 12. Questionnaire results of the in-situ evaluation of the mind alert based on internal thought detection.

Participants were required to complete a questionnaire of six questions: *Q1. You are interested in the mind alert. Q2. The audio alert is annoying. Q3. Your focus was enhanced. Q4. The frequency of mind wandering was* *reduced. Q5. Your awareness of mind wandering was enhanced. Q6. Your mind wandered prior to the alert.* All questions were rated on a 7-point Likert scale from "strongly disagree" (1) to "strongly agree" (7). The results are given in Fig 12. An interview was conducted afterward.

### 7.1. Results and Discussion

First, participants agreed that this mind alert was highly interesting (average score 6.6). Participant 2 (P2) mentioned, "*This is the thing I have been looking for!*" Interestingly, they also agreed that the alert, even the false ones, were not annoying (average score 1.4). This corroborates our claim that the audio alert only causes a negligible amount of cognitive load and requires no user intervention, and therefore did not have a strong negative impact on the users.

It is also encouraging to see that participants generally agreed that the mind alert enhanced their focus state (score=5.4), reduced the frequency of mind wandering (score=5.2), and increased their awareness of mind wandering (score=5.5). In addition, we received some promising comments, for instance, P1 reported, "*I was quite surprised that the alert came the moment I switched from my paper to Facebook*". P5 said, "*It's cool, like some guy is working with you and he knows your thinking*".

However, we see a limitation of training only on the lecture viewing data and testing in the wild. P2 complained it was oversensitive, "*... forced me to stare hard at the screen all the time when I watched a movie*". In contrast, P3 pointed out "*There were a few times that I realized I was mind wandering, but it didn't warn me… I was programming*". These issues may be due to task differences. Although vergence features can generalize well across tasks with similar visual attention, other tasks essentially require a higher visual focus (e.g. writing) while others demand more peripheral vision or saccadic movements (e.g. watching an action movie). The contextual information (Bulling & Zander, 2014), such as task and user intention, may influence eye vergence to a certain degree. However, we believe that this issue can be alleviated with a large-scale study and a diverse training dataset.

*Q5* is of the most interest to us. Participants gave mixed opinions with an overall score of 4.2. P5 said "*It gave a reasonable accuracy. I love to keep using it*". However, some comments exposed the limitation of the eye tracker. P4 said, "*The alert fired too much, especially when I looked near the edge of the monitor*". This may be due to the fact that the accuracy of the eye tracker decreased as the gaze moves further away from the centre, where our training data does not cover. P4 also mentioned "*It fired all the time in the last 30 minutes. Maybe I was getting tired, but I am sure it fired even [though] I was focused*". We found out that it was mainly caused by the head deviation from the optimal scope of the eye tracker.

### 7.2. Summary

The in-situ evaluation indicates vergence-based detection of internal thought is a promising technique. In practice, accurate gaze estimation and a diverse training set can contribute to detection robustness.

## 8. Discussion

In this work, we were the first to exploit eye vergence behaviour for light-weight and moment-to-moment detection of internal thought. This technique showed promising results for goal-directed thought detection during a math-solving task experiment as well as for internal thought detection during lecture viewing and in-situ experiments. Our evaluations demonstrated that our method can outperform existing method using fixations, saccades, and blinks in both math-solving task and the natural lecture viewing experiments. Moreover, within-task leave-one-participant-out cross-evaluation and cross-task evaluations both indicate the learning effect from the data obtained from our annotation paradigm. The encouraging feedback from an in-situ evaluation further demonstrates the potential of the vergence-based detection technique.

To acquire fine-grained ground truth of internal thought and maintain the natural user behaviour in task, we further proposed a novel annotation paradigm that exploits human blur perception. For the thorough understanding, we conducted a human blur perception study to determine the parameters for our annotation paradigm. Compared with the traditional self-caught and probe-caught techniques, our annotation paradigm is able to yield a conservative occurrence period of internal thought and it requires less intrusive intervention from the users, allowing them to sustain attention to the main task. It therefore opens new opportunities for human behavioural and psychological studies on attention.

One important property that differentiates our method from the state of the art is that vergence features can be extracted from a very small time window (1s). As shown in our work, this allows for moment-to-moment detection of internal thought and thus enables fine-grained understanding of users' attentive state. If combined with visual learning material, e.g. the content of a lecture video, future e-learning interfaces could provide more precise and effective interventions to optimise knowledge acquisition and memory retention. With the rapid emergence of e-learning applications and the absence of human educators in teaching, there is an increasing demand of intelligent user interface that understand and adapt to individual users. Our method provides a light-weight solution to moment-to-moment detection of internal thought and thus paves the way for numerous attention-aware applications in intelligent tutoring systems and human-computer interaction in general.

Despite these promising findings, there are still some limitations that we aim to address in future work. First, to enable real-time attention-aware interfaces using only affordable devices and to allow for the deployment to real-world learning environments, the eye tracker (Tobii EyeX) we used in our study is a consumer-grade tracker the same as Hutt et al. (2016). This attempt alleviates the constraint of the expensive research-grade eye tracker as pointed out by Bixler and D'Mello (2016). Although EyeX gives a similar accuracy (offset from the true gaze point) as Tobii Pro X2-60 as well as a reasonable precision (Feit et al., 2017) and its firmware is optimized for real-time interaction (Tobii AB, 2018), its frame rate still limits the accurate detection of some saccadic behaviours. This might affect our comparison of using the "classic" eye movement features, but it is currently unavoidable as research-grade eye trackers are too expensive for home-use intelligent interfaces. Besides, since we posed no explicit constraint on the head movement in our experiment to maintain the natural user behaviour, this might lead to an accuracy drop of gaze estimation (Niehorster, Cornelissen, Holmqvist, Hooge, & Hessels, 2017). Second, fatigue might have influenced internal thought in the lecture viewing experiment. Such affective biases are difficult to eliminate but our further in-situ evaluation confirmed the effectiveness of the model trained from the lecture viewing data. This implies that biases in lecture viewing experiment may be modest, if any. Third, the in-situ experiment also exposed that vergence features depend on accurate gaze estimation. This issue could be addressed using implicit calibration approaches (Huang, Kwok, Ngai, Chan, and Leong (2016) Papoutsaki et al. (2016)). Fourth, we hope to further explore vergence behaviour by representation learning from a large-scale in-the-wild data. As Solé Puig et al., (2013) pointed out, vergence can be affected by covert spatial attention and we foresee that taking contextual information (Bulling & Zander, 2014) or activities information (Bulling, Ward, Gellersen, & Tröster, 2011; Steil & Bulling, 2015) into consideration can result in a better representation for the detection of internal thought. Last but not least, we formulate the detection of internal thought as a binary classification task the same as the prior mind wandering studies (D'Mello, 2016). However, our results reveal that recognition gets harder on the verge of focused attention. We postulate the degree of internal thought could be reflected by the disparity degree of eye vergence and we plan to conduct further investigation on this in future.

## 9. Conclusion

In this work, we proposed a method for user-independent, moment-to-moment detection of internal thought from eye vergence behaviour. We further proposed a novel experimental paradigm to identify the start and end of internal thought in a fine-grained manner. Extensive evaluations demonstrated the effectiveness of our method for spontaneous internal thought detection as well as the generalizability of the proposed vergence features. Moment-to-moment detection of internal thought allows for a fine-grained understanding of human mental activities. Our findings thus not only contribute to the fundamental study of mental states but also open new applications, e.g. in mental health. As such, we believe moment-to-moment detection of internal thought has significant potential to become an indispensable component in user experience research.

## References

Andrzejewska, M., & Stolińska, A. (2013). Comparing the Difficulty of Tasks Using Eye Tracking Combined with Subjective and Behavioural Criteria, *9*(3), 1–16. https://doi.org/10.16910/jemr.9.3.3

Benedek, M., Stoiser, R., Walcher, S., & Körner, C. (2017). Eye Behavior Associated with Internally versus Externally Directed Cognition. *Frontiers in Psychology*, *8*(June), 1–9. https://doi.org/10.3389/fpsyg.2017.01092

Bixler, R., & D'Mello, S. (2016). Automatic gaze-based user-independent detection of mind wandering during computerized reading. *User Modeling and User-Adapted Interaction*, *26*(1), 33–68. https://doi.org/10.1007/s11257-015-9167-1

Breiman, L., & Leo Breiman. (2001). Random Forests. *Machine Learning*, *45*(1), 5–32. https://doi.org/10.1023/A:1010933404324

Bulling, A. (2016). Pervasive Attentive User Interfaces. *Computer*, *49*(1), 94–98. https://doi.org/10.1109/MC.2016.32

Bulling, A., & Roggen, D. (2011). Recognition of visual memory recall processes using eye movement analysis. In *Proceedings of the 13th international conference on Ubiquitous computing - UbiComp '11* (p. 455). New York, New York, USA: ACM Press. https://doi.org/10.1145/2030112.2030172

Bulling, A., Ward, J. A., & Gellersen, H. (2012). Multimodal recognition of reading activity in transit using body-worn sensors. *ACM Transactions on Applied Perception*, *9*(1), 1–21. https://doi.org/10.1145/2134203.2134205

Bulling, A., & Zander, T. O. (2014). Cognition-Aware Computing. *IEEE Pervasive Computing*, *13*(3), 80–83. https://doi.org/10.1109/MPRV.2014.42

Casiez, G., Roussel, N., & Vogel, D. (2012). 1 € filter: a simple speed-based low-pass filter for noisy input in interactive systems. In *Proceedings of the 2012 ACM annual conference on Human Factors in Computing Systems - CHI '12* (p. 2527). New York, New York, USA: ACM Press. https://doi.org/10.1145/2207676.2208639

Cheng, S., Sun, Z., Sun, L., Yee, K., & Dey, A. K. (2015). Gaze-Based Annotations for Reading Comprehension. *Proceedings of the 33rd Annual ACM Conference on Human Factors in Computing Systems - CHI '15*, *1*, 1569–1572. https://doi.org/10.1145/2702123.2702271

Christoff, K., Irving, Z. C., Fox, K. C. R., Spreng, R. N., & Andrews-Hanna, J. R. (2016). Mind-wandering as spontaneous thought: a dynamic framework. *Nature Reviews Neuroscience*, *17*(11), 718–731. https://doi.org/10.1038/nrn.2016.113

Conati, C., Aleven, V., & Mitrovic, A. (2013). Eye-Tracking for Student Modelling in Intelligent Tutoring Systems. In *Design Recommendations for Intelligent Tutoring Systems* (pp. 227–236).

D'Mello, S. K. (2016). Giving Eyesight to the Blind: Towards Attention-Aware AIED. *International Journal of Artificial Intelligence in Education*, *26*(2), 645–659. https://doi.org/10.1007/s40593-016-0104-1

D'Mello, S., Olney, A., Williams, C., & Hays, P. (2012). Gaze tutor: A gaze-reactive intelligent tutoring system. *International Journal of Human Computer Studies*, *70*, 377–398. https://doi.org/10.1016/j.ijhcs.2012.01.004

Dingler, T. (2016). Cognition-Aware systems as mobile personal assistants. *UbiComp 2016 Adjunct - Proceedings of the 2016 ACM International Joint Conference on Pervasive and Ubiquitous Computing*, 1035–1040. https://doi.org/10.1145/2968219.2968565

Dixon, M. L., Fox, K. C. R., & Christoff, K. (2014). A framework for understanding the relationship between externally and internally directed cognition. *Neuropsychologia*, *62*, 321–330. https://doi.org/10.1016/j.neuropsychologia.2014.05.024

Engbert, R., & Kliegl, R. (2003). Microsaccades uncover the orientation of covert attention. *Vision Research*, *43*(9), 1035–1045. https://doi.org/10.1016/S0042-6989(03)00084-1

Faber, M., Bixler, R., & D'Mello, S. K. (2017). An automated behavioral measure of mind wandering during computerized reading. *Behavior Research Methods*. https://doi.org/10.3758/s13428-017-0857-y

Feit, A. M., Williams, S., Toledo, A., Paradiso, A., Kulkarni, H., Kane, S., & Morris, M. R. (2017). Toward Everyday Gaze Input: Accuracy and Precision of Eye Tracking and Implications for Design. In *Proceedings of the 2017 CHI Conference on Human Factors in Computing Systems - CHI '17* (pp. 1118–1130). New York, New York, USA: ACM Press. https://doi.org/10.1145/3025453.3025599

Foulsham, T., Farley, J., & Kingstone, A. (2013). Mind wandering in sentence reading: Decoupling the link between mind and eye. *Canadian Journal of Experimental Psychology/Revue Canadienne de Psychologie Expérimentale*, *67*(1), 51–59. https://doi.org/10.1037/a0030217

Frank, D. J., Nara, B., Zavagnin, M., Touron, D. R., & Kane, M. J. (2015). Validating older adults' reports of less mind-wandering: An examination of eye movements and dispositional influences. *Psychology and Aging*, *30*(2), 266–278. https://doi.org/10.1037/pag0000031

Franklin, M. S., Broadway, J. M., Mrazek, M. D., Smallwood, J., & Schooler, J. W. (2013). Window to the wandering mind: Pupillometry of spontaneous thought while reading. *The Quarterly Journal of Experimental Psychology*, *66*(12), 2289–2294. https://doi.org/10.1080/17470218.2013.858170

Goldberg, J. H., & Kotval, X. P. (1999). Computer interface evaluation using eye movements: methods and constructs. *International Journal of Industrial Ergonomics*, *24*(6), 631–645. https://doi.org/10.1016/S0169-8141(98)00068-7

Hansen, D. W., & Ji, Q. (2010). In the eye of the beholder: a survey of models for eyes and gaze. *IEEE Transactions on Pattern Analysis and Machine Intelligence*, *32*(3), 478–500. https://doi.org/10.1109/TPAMI.2009.30

Hoppe, S., Loetscher, T., Morey, S., & Bulling, A. (2015). Recognition of curiosity using eye movement analysis. In *Proceedings of the 2015 ACM International Joint Conference on Pervasive and Ubiquitous Computing and Proceedings of the 2015 ACM International Symposium on Wearable Computers - UbiComp '15* (pp. 185–188). New York, New York, USA: ACM Press. https://doi.org/10.1145/2800835.2800910

Huang, M. X., Kwok, T. C. K., Ngai, G., Chan, S. C. F., & Leong, H. V. (2016). Building a Personalized, Auto-Calibrating Eye Tracker from User Interactions. In *Proceedings of the 2016 CHI Conference on Human Factors in Computing Systems* (pp. 5169–5179). https://doi.org/10.1145/2858036.2858404

Huang, M. X., Li, J., Ngai, G., & Leong, H. V. (2016). StressClick: Sensing Stress from Gaze-Click Patterns. In *Proceedings of ACM International Conference on Multimedia* (pp. 1395–1404).

Hutt, S., Hardey, J., Bixler, R., Stewart, A., Risko, E., & Mello, S. K. D. (2017). Gaze-based Detection of Mind Wandering during Lecture Viewing. In *10th International Conference on Educational Data Mining*.

Hutt, S., Mills, C., White, S., Donnelly, P. J., & D'Mello, S. K. (2016). The Eyes Have It: Gaze-based Detection of Mind Wandering during Learning with an Intelligent Tutoring System. In *Proceedings of the 9th International Conference on Educational Data Mining, International Educational Data Mining Society* (pp. 86–93).

Kardan, S., & Conati, C. (2015). Providing Adaptive Support in an Interactive Simulation for Learning. *Proceedings of the 33rd Annual ACM Conference on Human Factors in Computing Systems - CHI '15*, 3671–3680. https://doi.org/10.1145/2702123.2702424

Killingsworth, M. A., & Gilbert, D. T. (2010). A Wandering Mind Is an Unhappy Mind. *Science*, *330*(6006), 932–932. https://doi.org/10.1126/science.1192439

Kudo, S., Okabe, H., Hachisu, T., Sato, M., Fukushima, S., & Kajimoto, H. (2013). Input method using divergence eye movement. In *CHI '13 Extended Abstracts on Human Factors in Computing Systems on - CHI EA '13* (p. 1335). New York, New York, USA: ACM Press. https://doi.org/10.1145/2468356.2468594

Li, J., Ngai, G., Leong, H. V., & Chan, S. C. F. (2016). Your Eye Tells How Well You Comprehend. In *IEEE 40th Annual Computer Software and Applications Conference (COMPSAC)* (pp. 503–508). https://doi.org/10.1109/COMPSAC.2016.220

Mijović, P., Ković, V., De Vos, M., Mačužić, I., Todorović, P., Jeremić, B., & Gligorijević, I. (2017). Towards continuous and real-time attention monitoring at work: reaction time versus brain response. *Ergonomics*, *60*(2), 241–254. https://doi.org/10.1080/00140139.2016.1142121

Mills, C., Bixler, R., Wang, X., & Mello, S. K. D. (2016). Automatic Gaze-Based Detection of Mind Wandering during Narrative Film Comprehension. In *Proceedings of the 9th International Conference on Educational Data Mining (EDM 2016)* (pp. 30–37).

Mishra, A., Kanojia, D., Nagar, S., Dey, K., & Bhattacharyya, P. (2017). Scanpath Complexity : Modeling Reading Effort Using Gaze Information. *Proceedings of the 31th Conference on Artificial Intelligence (AAAI 2017)*, 4429–4436.

Navalpakkam, V., Kumar, R., Li, L., & Sivakumar, D. (2012). Attention and Selection in Online Choice Tasks. In *International Conference on User Modeling, Adaptation, and Personalization* (pp. 200–211). https://doi.org/10.1007/978-3-642-31454-4_17

Niehorster, D. C., Cornelissen, T. H. W., Holmqvist, K., Hooge, I. T. C., & Hessels, R. S. (2017). What to expect from your remote eye-tracker when participants are unrestrained. *Behavior Research Methods*. https://doi.org/10.3758/s13428-017-0863-0

Oh, J., Jeong, S.-Y., & Jeong, J. (2012). The timing and temporal patterns of eye blinking are dynamically modulated by attention. *Human Movement Science*, *31*(6), 1353–1365. https://doi.org/10.1016/j.humov.2012.06.003

Olney, A. M., Risko, E. F., D'Mello, S. K., & Graesser, A. C. (2015). Attention in Educational Contexts: The Role of the Learning Task in Guiding Attention. In *The Handbook of Attention* (pp. 623–642). MIT Press.

Papoutsaki, A., Sangkloy, P., Laskey, J., Daskalova, N., Huang, J., & Hays, J. (2016). WebGazer: Scalable Webcam Eye Tracking Using User Interactions. In *Proceedings of the 25th International Joint Conference on Artificial Intelligence (IJCAI)* (pp. 3839–3845).

Pfleging, B., Fekety, D. K., Schmidt, A., & Kun, A. L. (2016). A Model Relating Pupil Diameter to Mental Workload and Lighting Conditions. *CHI Conference on Human Factors in Computing Systems*, 5776–5788. https://doi.org/10.1145/2858036.2858117

Pham, P., & Wang, J. (2015). AttentiveLearner: Improving Mobile MOOC Learning via Implicit Heart Rate Tracking. In *International Conference on Artificial Intelligence in Education* (Vol. 9112, pp. 367–376). https://doi.org/10.1007/978-3-319-19773-9_37

Rayner, K. (1998). Eye movements in Reading and Information Processing: 20 Years of Research. *Psychological Bulletin*, *124*(3), 372–422. https://doi.org/10.1037/0033-2909.124.3.372

Reichle, E. D., Reineberg, A. E., & Schooler, J. W. (2010). Eye Movements During Mindless Reading. *Psychological Science*, *21*, 1300–1310. https://doi.org/10.1109/ICCVW.2017.322

Salvucci, D. D., & Goldberg, J. H. (2000). Identifying fixations and saccades in eye-tracking protocols. In *Proceedings of the symposium on Eye tracking research & applications - ETRA '00* (pp. 71–78). New York, New York, USA, New York, USA: ACM Press. https://doi.org/10.1145/355017.355028

Sattar, H., Bulling, A., & Fritz, M. (2017). Predicting the Category and Attributes of Visual Search Targets Using Deep Gaze Pooling. In *2017 IEEE International Conference on Computer Vision Workshops (ICCVW)* (pp. 2740–2748). IEEE. https://doi.org/10.1109/ICCVW.2017.322

Sattar, H., Muller, S., Fritz, M., & Bulling, A. (2015). Prediction of search targets from fixations in open-world settings. In *2015 IEEE Conference on Computer Vision and Pattern Recognition (CVPR)* (pp. 981–990). IEEE. https://doi.org/10.1109/CVPR.2015.7298700

Schooler, J. W., Smallwood, J., Christoff, K., Handy, T. C., Reichle, E. D., & Sayette, M. A. (2011). Meta-awareness, perceptual decoupling and the wandering mind. *Trends in Cognitive Sciences*. https://doi.org/10.1016/j.tics.2011.05.006

Smallwood, J., Davies, J. B., Heim, D., Finnigan, F., Sudberry, M., O'Connor, R., & Obonsawin, M. (2004). Subjective experience and the attentional lapse: Task engagement and disengagement during sustained attention. *Consciousness and Cognition*, *13*(4), 657–690. https://doi.org/10.1016/j.concog.2004.06.003

Smallwood, J., O'Connor, R. C., Sudberry, M. V., Haskell, C., & Ballantyne, C. (2004). The consequences of encoding information on the maintenance of internally generated images and thoughts: The role of meaning complexes. *Consciousness and Cognition*, *13*(4), 789–820. https://doi.org/10.1016/j.concog.2004.07.004

Smallwood, J., & Schooler, J. W. (2006). The restless mind. *Psychological Bulletin*, *132*(6), 946–958. https://doi.org/10.1037/0033-2909.132.6.946

Smallwood, J., & Schooler, J. W. (2015). The Science of Mind Wandering: Empirically Navigating the Stream of Consciousness. *Annual Review of Psychology*, *66*(1), 487–518. https://doi.org/10.1146/annurev-psych-010814-015331

Smilek, D., Carriere, J. S. A., & Cheyne, J. A. (2010). Out of Mind, Out of Sight: eye blinking as indicator and embodiment of mind wandering. *Psychological Science*, *21*(6), 786–789. https://doi.org/10.1177/0956797610368063

Solé Puig, M., Pérez Zapata, L., Aznar-Casanova, J. A., & Supèr, H. (2013). A Role of Eye Vergence in Covert Attention. *PLoS ONE*, *8*(1), e52955. https://doi.org/10.1371/journal.pone.0052955

Solé Puig, M., Pérez Zapata, L., Puigcerver, L., Esperalba Iglesias, N., Sanchez Garcia, C., Romeo, A., … Supèr, H. (2015). Attention-Related Eye Vergence Measured in Children with Attention Deficit Hyperactivity Disorder. *PLOS ONE*, *10*(12), e0145281. https://doi.org/10.1371/journal.pone.0145281

Steichen, B., Wu, M. M. a, Toker, D., Conati, C., & Carenini, G. (2014). Te,Te,Hi,Hi: Eye gaze sequence analysis for informing user-adaptive information visualizations. *Lecture Notes in Computer Science (Including Subseries Lecture Notes in Artificial Intelligence and Lecture Notes in Bioinformatics)*, *8538*, 183–194. https://doi.org/10.1007/978-3-319-08786-3_16

Steil, J., & Bulling, A. (2015). Discovery of everyday human activities from long-term visual behaviour using topic models. In *Proceedings of the 2015 ACM International Joint Conference on Pervasive and Ubiquitous Computing - UbiComp '15* (pp. 75–85). New York, New York, USA: ACM Press.

https://doi.org/10.1145/2750858.2807520

Toates, F. M. (1974). Vergence eye movements. *Documenta Ophthalmologica*, *37*(1), 153–214. https://doi.org/10.1007/BF00149678

Tobii AB. (2018). http://developer.tobii.com/.

Toker, D., & Conati, C. (2017). Leveraging Pupil Dilation Measures for Understanding Users' Cognitive Load During Visualization Processing. In *Adjunct Publication of the 25th Conference on User Modeling, Adaptation and Personalization - UMAP '17* (pp. 267–270). New York, New York, USA: ACM Press. https://doi.org/10.1145/3099023.3099059

Unsworth, N., & Robison, M. K. (2016). Pupillary correlates of lapses of sustained attention. *Cognitive, Affective, & Behavioral Neuroscience*, (April), 601–615. https://doi.org/10.3758/s13415-016-0417-4

Varela Casal, P., Lorena Esposito, F., Morata Martínez, I., Capdevila, A., Solé Puig, M., de la Osa, N., … Cañete, J. (2018). Clinical Validation of Eye Vergence as an Objective Marker for Diagnosis of ADHD in Children. *Journal of Attention Disorders*, 108705471774993. https://doi.org/10.1177/1087054717749931

Vertegaal, R. (2003). Attentive User Interfaces. *Communications of the ACM*, *46*(3), 30. https://doi.org/10.1145/636772.636794

Walcher, S., Körner, C., & Benedek, M. (2017). Looking for ideas: Eye behavior during goal-directed internally focused cognition. *Consciousness and Cognition*, *53*, 165–175. https://doi.org/10.1016/j.concog.2017.06.009

Watson, A. B., & Ahumada, A. J. (2011). Blur clarified: A review and synthesis of blur discrimination. *Journal of Vision*, *11*(5), 10–10. https://doi.org/10.1167/11.5.10

Wickens, C. D. (1981). *Processing Resources in Attention, Dual Task Performance, and Workload Assessment*. Defense Technical Information Center.

Xiao, X., & Wang, J. (2017). Undertanding and Detecting Divided Attention in Mobile MOOC Learning. *Proceedings of the 2017 CHI Conference on Human Factors in Computing Systems - CHI '17*, 2411–2415. https://doi.org/10.1145/3025453.3025552

Xiaohui Shen, & Ying Wu. (2012). A unified approach to salient object detection via low rank matrix recovery. In *2012 IEEE Conference on Computer Vision and Pattern Recognition* (pp. 853–860). IEEE. https://doi.org/10.1109/CVPR.2012.6247758